\documentclass[conference]{IEEEtran}
\usepackage{cite}
\ifCLASSINFOpdf
 
\else
\usepackage[dvips]{graphicx}
\fi
\usepackage[cmex10]{amsmath}

\IEEEoverridecommandlockouts \IEEEpubid{\makebox[\columnwidth]{ICNF2013\hfill 978-1-4799-0671-0/13/\$31.00~\copyright 2013~IEEE} \hspace{\columnsep}\makebox[\columnwidth]{ }}

\begin{document}
\def\Journal#1#2#3#4{{#1} {\bf #2,} { #3} (#4)} 
 
% Some useful journal names 
\def\BiJ{ Biophys. J.}                 
\def\Bios{ Biosensors and Bioelectronics} 
\def\LNC{ Lett. Nuovo Cimento} 
\def\FNL{ Fluct. Noise Lett.}
\def\JCP{ J. Chem. Phys.} 
\def\JAP{ J. Appl. Phys.}
\def\JACS{ J. Am. Chem. Soc.} 
\def\JMB{ J. Mol. Biol.} 
\def\CMP{ Comm. Math. Phys.} 
\def\LMP{ Lett. Math. Phys.} 
\def\NLE{{ Nature Lett.}} 
\def\NPB{{ Nucl. Phys.} B} 
\def\PLA{{ Phys. Lett.}  A} 
\def\PLB{{ Phys. Lett.}  B} 
\def\PRL{ Phys. Rev. Lett.} 
\def\PRA{{ Phys. Rev.} A} 
\def\PRE{{ Phys. Rev.} E} 
\def\PRB{{ Phys. Rev.} B} 
\def\PNAS{{ Proc. Nat. Acad. Sci.}}
\def\EPL{{Europhys. Lett.} } 
\def\PD{{ Physica} D} 
\def\ZPC{{ Z. Phys.} C} 
\def\RMP{ Rev. Mod. Phys.} 
\def\EPJD{{ Eur. Phys. J.} D} 
\def\SAB{ Sens. Act. B} 
\title{Current-Voltage Characteristics and non-Gaussian fluctuations in two different protein light receptors}

\author{\IEEEauthorblockN{E. Alfinito}
\IEEEauthorblockA{Dipartimento di Ingegneria dell'Innovazione \\
 Universit\`a del Salento, via Monteroni, I-73100 Lecce-Italy\\
 and CNISM,  via della Vasca Navale,84 I-00146 Roma-Italy\\
Email: eleonora.alfinito@unisalento.it}\and
\IEEEauthorblockN{J. Pousset and L. Reggiani}
\IEEEauthorblockA{Dipartimento di Matematica e Fisica "Ennio De Giorgi"\\ Universit\'a del Salento Via Monteroni, I-73100 Lecce, Italy
}}

\maketitle

\begin{abstract}
We investigate conductance and conductance fluctuations of two transmembrane proteins, 
bacteriorhodopsin and proteorhodopsin, belonging to the family of protein light receptors.
These proteins are widely diffused in aqueous environments, are sensitive to visible light and are promising biomaterials
for the realization of novel photodevices.
The conductance exhibits a rapid increase at increasing applied voltages, over a threshold value. Around the threshold value the variance of conductance
fluctuations
shows a dramatic jump of about 5 orders of magnitude: conductance and variance behaviours trace a second order phase transition.
Furthermore, the conductance fluctuations evidence a non-Gaussian behaviour with a probability density function (PDF) which follows a generalized Gumbel distribution, typical of extreme-value statistics. 
The theoretical model is validated on existing current-voltage measurements  and the interpretation of the PDF of conductance fluctuations is proven to be in line with the microscopic mechanisms responsible of charge transport. 
\end{abstract}


\date{\today}

% \fi
\IEEEpeerreviewmaketitle

\section{Introduction}
Recent advances in biological materials, to be used as frontier matter
for the development of nano-biodevices, open new questions about the possible ways to monitor their properties under stressing conditions. 
In particular, when the stress has an electrical origin, it is known that a proper tool of investigation is given by the study of electrical fluctuations, that are more sensitive to characterize the system degradation than the associated average quantities. 
A couple of proteins, naturally working as light receptors, bacteriorhodopsin
(bR), found in the Archean microorganism \textit{Halobacterium salinarum} \cite{Oerst}, and proteorhodopsin (pR), found in an uncultivated marine bacterium (SAR-86group) \cite{Beja},
have been recently proposed as active matter when characterized by current voltage (I-V) measurements in  metal-protein-metal (MPM) structures \cite{Jin,
Lee}. 
The experiments were found of relevant interest since both materials  showed
a behaviour quite similar to that of a standard semiconductor with a low conductivity, of the order of $10^{-7} \ S/cm$, which 
significantly increases in the presence of a visible light and a stressing voltage. 
In particular, bR in forms of nanometric monolayers was monitored on a wide range of applied electric fields
(up to $10^4 \ kV/cm$) where two distinct regimes of response, a direct-tunneling (DT) and a Fowler-Nordheim (FN) tunneling regime were observed \cite{Gomila}.
By contrast, pR in form of thin film structures with an active region of about 50  micron length was explored 
on a limited range of applied electric fields
(up to $30 \ kV/cm$), where a substantial Ohmic behaviour was found \cite{Lee}.
\par 
Starting from available I-V investigations, the aim of this paper is to model
 the corresponding conductance and conductance fluctuations in a wide range of applied bias (7 orders of magnitude). In doing so, we follow a 
theoretical/computational approach, hereafter called INPA (impedance network
protein analogue), 
able to produce, for an assigned protein configuration,
the instantaneous chord conductance $g=I/V$ and the related fluctuations.  The average chord conductance, $<g>$, is compared with the experimental
data, as function of the applied voltage.
Furthermore, conductance fluctuations around the average value, obtained from the theoretical simulations, are analyzed in terms of the probability density functions (PDFs).
Interesting enough, the analogous of a phase transition of second order is evidenced in the conductance vs voltage behaviour near to the crossing between DT and FN tunneling regimes.
In particular, the PDF is found to exhibit a generalized Gumbel behaviour typical of  extreme-value statistics describing physical phenomena of very different nature \cite{Noullez,Bertin,Ciliberto,pre2002}.  
Accordingly, present results add a new example of phase transition where the 
generalized Gumbel distribution is recovered.
%%%%%%%%%%%%%%%%%%%%%%%%%%%%%%%%%%%%%%%%%%%%%%%%%%%%%%%%%%%%%%%%%%%%%%%%%%%%5
\section{Theory}
The measurements of I-V characteristics  performed in several proteins
\cite{Gomila,
Lee, Jin} provided the clear message that the protein morphology has a preeminent
role in the electrical response. 
In particular, it has been observed that
the conformational change associated with the protein light-activation, considerably increases the amount of current the protein is able to sustain
at a given voltage. 
This observation  suggests to overcome the existing theoretical approaches \cite{Gomila,Wang}: as matter of fact, they only provide a phenomenological interpretation of some I-V characteristics in terms of MPM, i.e. they appear as \textsl{macroscopic} tunneling theories. 
By contrast, the impedance network protein analogue model used here  correlates electrical and topological protein properties in such a way to predict how the modifications of the latter induce modifications in the former.
In brief, the INPA model \cite{PRE} maps the single protein into an electrical network, whose nodes correspond to the amino acids and the links between nodes are elementary resistances; in this way the current response is a consequence of the protein topology. 
The network is electrically solved using Kirchhoff's laws and an iterative approach of Monte Carlo (MC) type.
Accordingly, the values of the link resistances are stochastically chosen between a high resistivity value, $\rho(V)$, and a low value, $\rho_{min}$,  by means of a transmission probability whose voltage dependence interpolates between DT and FN tunneling. 
The high resistivity value is taken as:
\begin{equation}
\rho(V)=\left\{\begin{array}{lll}
\rho_{MAX}& \hspace{.0cm }& eV \le  \Phi  \\ \\
 \rho_{MAX} (\frac{\Phi}{eV})+\rho_{min}(1- \frac{\Phi}{eV}) &\hspace{.0cm} & eV \ge  \Phi 
 \end{array}
  \right.
\label{eq:3}
\end{equation}
where $\rho_{MAX}$ is the highest resistivity value used to fit the I-V characteristic at the lowest voltages (Ohmic response),  $\rho_{min} \ll \rho_{MAX}$  is the lowest resistive value used to fit the current at the highest voltages,  and $\Phi$  is the height of the tunneling barrier between nodes.
%
%The above interpolated formula reflects the different voltage dependence %in the pre-factor of the current expression \cite{Wang}: $I\sim V$ in the %DT regime, and $I\sim V^2$ in the FN tunneling regime.
\par
Due to the protein size, about 5 nm  in diameter, the charge transfer is here interpreted in terms of a \textit{sequential} tunneling between neighbouring amino acids and not, as usually found in literature \cite{Wang,Gomila}, in terms of a single tunneling between the metallic contacts. 
Here, two amino acids are considered neighbouring when they are inside a
\textit{sphere of interaction}, of radius $R_{C}$ \cite{PRE}.
The transmission probability of each tunneling process takes the form:
\begin{equation} 
\mathcal{P}^{D}_{i,j}= \exp \left[- \frac{2 l_{i,j}}{\hbar} \sqrt{2m(\Phi-\frac{1}{2}
eV_{i,j})} \right] \ 
\hspace{0.1cm}
 eV_{i,j}  \le \Phi \
\label{eq:1}
\end{equation}
\begin{equation}
\mathcal{P}^{FN}_{ij}=\exp \left[-\left(\frac{2l_{i,j}\sqrt{2m}}{\hbar}\right)\frac{\Phi}{eV_{i,j}}\sqrt{\frac{\Phi}{2}} \right] \  
\hspace{0.1cm}
 eV_{i,j} \ge \Phi \ 
\label{eq:2}
\end{equation}
where $V_{i,j}$ is the potential drop between the couple of $i,j$ amino-acids far apart for a distance  $l_{i,j}$, $\hbar$ is the reduced Planck constant, and $m$  is the electron effective mass, here taken the same of the bare value.
%\par
%The two transmission probabilities (Eqs.(\ref{eq:1},\ref{eq:2})), are plotted %in the inset of fig.~\ref{fig:1}, as function of the voltage drop between %the couple of nodes $i,j$. 
%We observe that these probabilities overlap in a bias region delimited by % the critical value, $V_C$, and the Ginzburg equivalent value, $V_G$ \cite{PRB2002,FNL}.
%These two values signal, respectively, the starting and the ending of the %\textit{nucleation/inertial} region \cite{Fisher,Bramwell2000}, in which %the FN regime permanently consolidates. 
%\par
%In order to identify the macroscopic analogue of $V_C$ and $V_G$, we calculate %the average conductance $<g>$, taken as the measurable  global-quantity.
%Figure~\ref{fig:1} reports the comparison between experiments and calculations %where the model.
\par
By construction, the iterative calculation of the conductance provides the average value as the mean value over the sequence of stochastic configurations compatible with the given applied voltage, as well as the fluctuations around the average value.  
Finally, being INPA a single-protein model, the results obtained should be appropriately scaled to be compared with the macroscopic experimental results. 
In the present case we use: $R_{C}=6$~{\AA}, $\rho_{MAX}= 4 \times 10^{13}{\Omega}$~\AA, and $\rho_{min}= 4 \times 10^{5}\Omega $~\AA, which produce a reasonable good agreement with the conductance measurements carried out on the considered opsins \cite{PRE}. 
\section{Results and discussion}
To compare bR and pR we start by choosing the
protein tertiary structures as the 2NTU protein-data-base entry \cite{PDB} for bR and
the 2L6X (model 13) entry for pR. 
Both entries refer to the protein structure in dark, i.e. in the native state.
\subsection{Average conductance and its variance}
The iterative evolution of the conductance is calculated for both proteins. 
\begin{figure}[!t]
\centering
\includegraphics[width=3.0in]{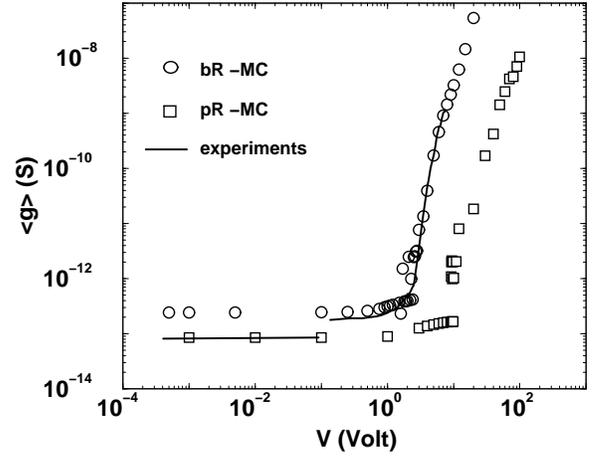}
\caption{Average conductance vs applied voltage for bR (full circles) and pR (empty squares). Values refer to those of a single protein.}
\label{fig_1}
\end{figure}
The  voltage dependence of the average conductance for both proteins is reported in Fig. \ref{fig_1}.
Here, an initial region of constant value is followed by a moderate increase (super-Ohmic region driven by the DT regime) which concludes with a very rapid growth above the threshold values (about 2.5 and 10 Volts,
respectively). 
The sharp growth is associated with the cross-over between DT and FN regimes.
In the case of bR, the Ohmic value of average conductance is greater than that of pR for about a factor of two and, as a consequence,  the predominance of the FN regime is definitely established at bias value of about a factor of four lower than that observed in pR.
We remark the good agreement of theoretical calculations with available experiments, which for the case of bR extends to the whole regime of biases considered.
\begin{figure}[!t]
\centering
\includegraphics[width=3.0in]{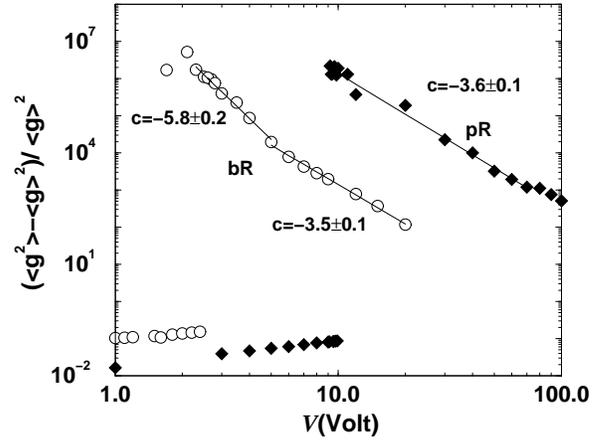}
\caption{Variance of conductance vs applied voltage
for bR (empty circles) and pR (full diamonds). The straight lines are the fitted
scaling laws (see text).}
\label{fig_2a}
\end{figure}
Figure ~\ref{fig_2a} reports the variance of conductance for both the proteins. The most relevant feature of this figure is the jump of about 5 orders of magnitude, close to the transition between the direct and FN regime. Above the transition it is possible to evidence a power law behavior of variance, with a single exponent close to 3 for bR and two different values for bR. As shown in the following, the different values of the critical exponent for bR suggest to search for a rescaling of the independent variable $V$.
\begin{figure}[!t]
\centering
\includegraphics[width=3.0in]{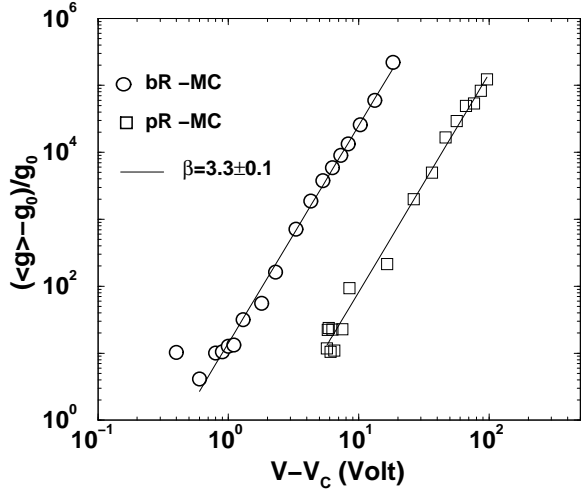}
\caption{Excess conductance vs scaled applied voltage
for bR (full circles) and pR (empty circles). The straight lines are the fitted
scaling laws (see text); $\beta$ is the critical exponent.}
\label{fig_2}
\end{figure}
\par
Figure~\ref{fig_2} reports the excess conductance 
g$_{EX}$=$(<g>-g_{0})/g_{0}$ vs the scaled applied bias, $V-V_{C}$.
Here $g_{0} = 2.43 \times 10^{-13} \ {S}$ and
$g_{0} = 8.54 \times 10^{-14} \ {S}$ indicate the low field average conductance  of bR (circles) and pR (squares), respectively. 
The value of $V_{C}$ is 1.7 ~V for bR and 3.5 ~V for pR, respectively.
Remarkably, g$_{EX}$  is found to follow the  power law:
\begin{equation}
g_{EX} \propto (V-V_C)^{\beta} 
\label{eq:pl1}
\end{equation}
with the critical exponent $\beta=3.3$ for both the proteins,
as reported in Fig.~\ref{fig_2}.
\begin{figure}[!t]
\centering
\includegraphics[width=3.0in]{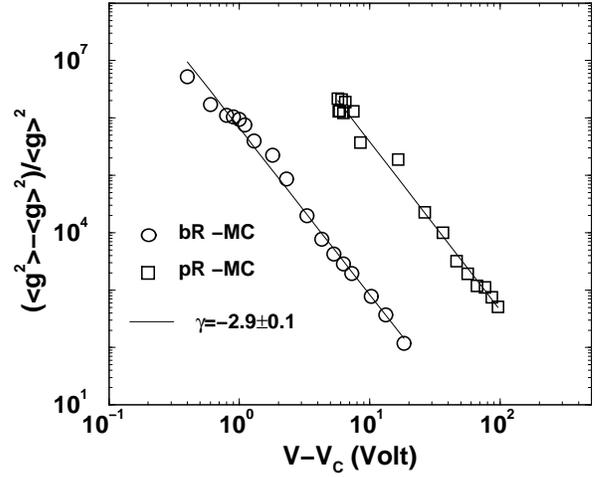}
\caption{Normalized variance of conductance fluctuations vs applied voltage for bR (full circles) and pR (empty squares), respectively. 
The straight lines are the fitted scaling laws (see text); $\gamma$ is
the critical exponent.
}
\label{fig_3}
\end{figure}
\par
Interesting enough, the normalized variance of conductance fluctuations, $S_{g}=(<g^{2}>-<g>^{2})/<g>^{2}$  
also evidences a power law behaviour
\begin{equation}
S_{g} \propto (V-V_{C})^{\gamma}
\end{equation} 
with the critical exponent $\gamma=-2.90$ for both the proteins, as reported in Fig.~\ref{fig_3}. 
We notice that the value of the critical exponent $\beta$  is very close to that of to thermal excess-conductance in superconductive thin films, in the regime of short-wave fluctuations, where it was found $\beta=3.2$ \cite{Aswal}, and that of resistance fluctuations in percolative systems, where it was found  $\beta=3.7$ \cite{pre2002}.
Therefore, we conjecture that all these behaviours share a common feature, mainly due to the granular nature of the samples. 
As a matter of fact, in the case of superconductivity \cite{Aswal}, the exponent value is related to the grains inside YBCO thin films, while in percolation models a similar behaviour is associated with the formation of clusters \cite{Stauffer}. 
We conclude that the  voltage-dependence of the conductance and of the variance of its fluctuations parallel those of a  second-order phase transition typical of finite-size systems \cite{Bramwell2000}.
\subsection{Probability distribution functions}
To support the conclusion of the  previous section,  we calculate the probability distribution functions (PDFs) of conductance  fluctuations, for both the proteins. 
By using a standard procedure of normalization \cite{Noullez}, it is possible to identify the PDF that best fits the simulations  as that belonging to the class of the generalized Gumbel distribution, $G(a)$ \cite{Noullez,Bertin}: 
\begin{equation}
G(a)=\frac{\theta(a)a^{a}}{\Gamma(a)}exp\{-aw-ae^{-w}\}
\label{eq:a1}
\end{equation}
%\begin{equation}
%G(a)=\frac{\theta(a)a^{a}}{\Gamma(a)}\exp\{-a[\theta(a)(z+\nu(a))+e^{-\theta(a)(z+\nu(a))}]\}
%\label{eq:ga}
%\end{equation}
%
where $a$ is a positive numerical parameter, and $w$ is the real variable: 
%the variable $z$ is the scaled conductance :
$w=\theta(a)(z+\nu(a))$ with
$z=(ln(g)-<ln(g)>)/\sigma_{a}$,   $\sigma_{a}$ being the standard deviation.
The function $\nu(a)$ is defined in terms of the Gamma function $\Gamma(a)$ and its derivatives
$\nu(a) ={1}/{\theta(a)}\left(\ln(a)-\psi(a)\right)$, $\Gamma(a)$, $\psi(a)$ and $\theta^{2}(a)$ indicating, respectively, the Gamma, digamma and trigamma function.
Finally, $G(a)$ is a normalized distribution function with zero mean and unitary variance.
\par
\begin{figure}[!t]
\centering
\includegraphics[width=3.0in]{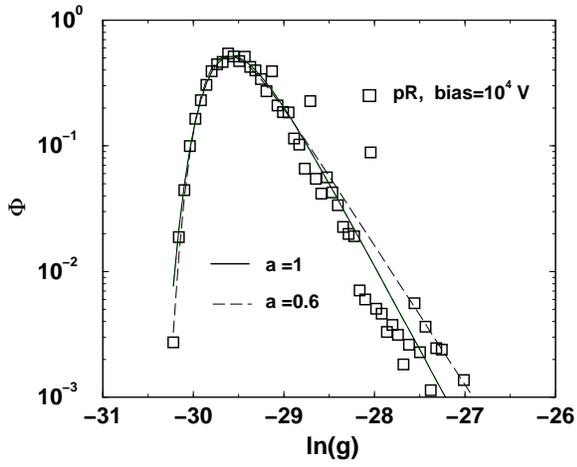}
\caption{Fitting functions of the pR conductance fluctuations at very high bias ($10^{4}V$). Both the fitting functions are Generalized Gumbel distribution without normalization \cite{EPL}. }
\label{fig_5}
\end{figure}
\begin{figure}[!t]
\centering
\includegraphics[width=3.0in]{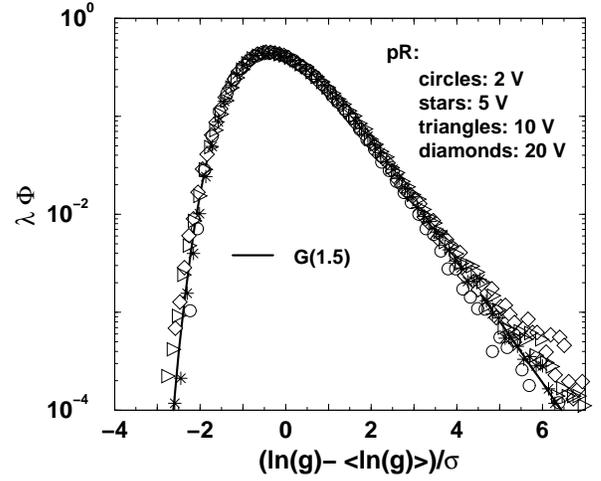}
\caption{PDFs of conductance fluctuations for pR. Continuous line indicates the
G(1.5) distributions, symbols the calculated data: circles for 2~V, stars for 5~V, triangles for 10~V, diamonds
for 20~V.}
\label{fig_6}
\end{figure}
\begin{figure}[!t]
\centering
\includegraphics[width=3.0in]{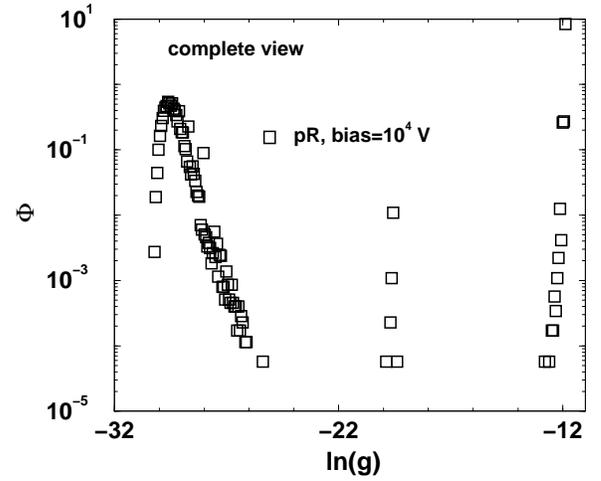}
\caption{Complete view of PDFs of conductance fluctuations for the case of pR at the highest bias of $10^{4} \ V$. 
}
\label{fig_7}
\end{figure}

Looking at the average quantities, it should be expected that also the PDFs
of bR and pR conductance fluctuations would share a very similar behaviour. 
However, this is not completely the case. 
\begin{figure}[!t]
\centering
\includegraphics[width=3.0in]{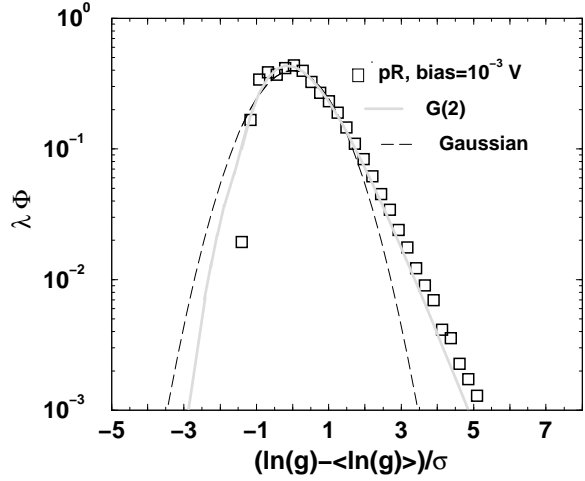}
\caption{PDFs of conductance fluctuations for pR at very low bias ($10^{-3}V$). Symbols refer to calculated data, continuous line is the $G(2)$ fit and dashed line is the Gaussian fit.  }
\label{fig_4}
\end{figure}
From one hand, and as a common trait with bR, the value of the shaping  parameter $a$ pertaining to the pR PDFs decreases at increasing bias, from $a=2$, see Fig.~\ref{fig_4}, to values of $a$ lower than 1 at very high
bias, as reported in Fig.~\ref{fig_5}. 
From another hand, when compared with bR, at low bias pR does not exhibit  a bimodal shape \cite{EPL}. 
In the intermediate bias region, as previously found in bR \cite{EPL,FNL}, the pR PDFs share a common $a$ value, $a=1.5$, as reported in  Fig.~\ref{fig_6},
while it is equal to 1.0 in bR \cite{EPL}. 
Finally, at the highest bias (above about 20 V) the PDFs exhibit very sharp and distant  peaks, as reported in Fig.~\ref{fig_7}, and this is again a
common trait with the case of  bR.
\par
In bacteriorhodopsin, the bias region that pertains to the $a=1$ value of the $G(a)$ distributions, signals the presence of an instability region in which DT and FN tunneling regimes compete \cite{EPL}. 
This region starts at about 1 V in bR, i.e. quite before the average conductance shows instability. 
A similar behaviour is observed in the case of pR. 
%%%%%%%%%%%%%%%%%%%%%%%%%%%%%%%%

% As probably could be expected, the high conductance peak is symmetric and %corresponds to the high bias Ohmic regime.
%%
%
%
\section{Conclusion}
The paper investigates the conductance and its fluctuations at increasing applied voltages in MPM structures where the active region consists of light-activated protein receptors, expressed in nature in different bacteria. 
These proteins  exhibit a very similar global morphology, i.e. they are seven helix transmembrane receptors. 
A detailed comparison between structures made with two of these proteins, bR and pR, has been carried out by using a theoretical microscopic  model based on an impedance network protein analogue.
Calculations compare well with available experiments, and offer challenging opportunities for further developments. 
The main results obtained can be summarized as follows:
the voltage dependent conductance of a single protein exhibits a second order phase transition at applied bias of a few Volts. 
This critical behaviour is associated with the cross-over from a charge transport controlled by a direct tunneling mechanism to that controlled  by a Fowler-Nordheim tunneling mechanism.
The phase transition is confirmed by a predicted sharp (over five orders of magnitude\cite{EPL}) increase of the variance of conductance fluctuations  and by the existence of power laws that fit the conductance and its fluctuations vs applied voltage.
The PDFs of conductance fluctuations are found to be strongly nongaussian following the shape of a generalized Gumbel distribution $G(a)$ and they show significative differences when going from bR to pR, which are associated with the different protein morphology.  
In general, the shape parameter $a$ is a function of the applied bias.
Despite being a first step, we believe that the electrical properties of the opsins here analyzed  open a scenario of relevant  interest for other 
biomaterials  belonging to the same family of transmembrane proteins.

\section*{Acknowledgment} 
%\begin{acknowledgments}
%
%\par
%\ack
Research supported by the European Commission under the  Bioelectronic Olfactory Neuron Device (BOND) project sponsored within grant agreement number 228685-2.
%
%\end{acknowledgments}
\bibliographystyle{IEEEtran}

\end{document}